\documentclass[twocolumn,amsmath,amssymb, prb]{revtex4}
\usepackage{graphicx}
\usepackage{latexsym}
\usepackage{amssymb}
\usepackage{amsmath}

\begin{document}

\title{The Diamond SQUID}

\author{Soumen Mandal$^{1,\footnote[1]{Corresponding authors: soumen.mandal@gmail.com, bauerle@grenoble.cnrs.fr}}$}
\author{Tobias Bautze$^1$}
\author{Oliver A. Williams$^2$}
\author{C\'ecile Naud$^1$}
\author{\'{E}tienne Bustarret$^1$}
\author{Franck Omn\`es$^1$}
\author{Pierre Rodi\`ere$^1$}
\author{Tristan Meunier$^1$}
\author{Christopher B\"{a}uerle$^{1,\footnotemark[1]}$}
\author{Laurent Saminadayar$^{1,3}$}
\affiliation{$^1$Institut N\'{e}el, CNRS et Universit\'{e} Joseph Fourier, F-38042 Grenoble, France}
\affiliation{$^2$Fraunhofer-Institut f\"ur Angewandte Festk\"{o}rperphysik, Tullastra{\ss}e 72, 79108 Freiburg, Germany}
\affiliation{$^3$Institut Universitaire de France, 103 boulevard Saint-Michel, 75005 Paris, France}

\begin{abstract}
Diamond is an electrical insulator in its natural
form. However, when doped with boron above a critical level ($\sim$
0.25 at.\%) it can be rendered superconducting at low temperatures
with high critical fields. Here we present the realization of a
micrometer scale superconducting quantum interference device
($\mu$-SQUID) made from nanocrystalline boron doped diamond (BDD)
films. Our results demonstrate that $\mu$-SQUIDs made from
superconducting diamond can be operated in magnetic fields as
large as 4T independent on the field direction. This is a decisive step
towards the detection of quantum motion in a diamond based nanomechanical oscillator.
\end{abstract}

\maketitle

Micro and nano SQUIDS\cite{01Cleziou} are extremely sensitive
tools for magnetization measurements on the local scale and find
applications in various fields of science such as scanning SQUID
microscopy\cite{02Kirtley}. However, the present state of the art
limits its utility to magnetic fields well below a tesla. Even though,
several attempts to realize micrometer scale superconducting quantum interference
devices ($\mu$-SQUIDs) form materials with high critical field such as
Nb$_3$Sn \cite{2aWu}  or Nb$_3$Ge \cite{2bDilorio,2cRogalla}  have been realized,
the demonstration of a device remaining operational at high magnetic fields has been elusive to date.
In this context, diamond, when doped with boron
above a critical level ($\sim$ 0.25 at.\%) which results in a superconductor with
very high critical field\cite{07Takano} is an extremely promising material. In
addition, recent advances in diamond thin film growth technology have paved the
way toward large scale processing of high quality devices.This recently discovered
material not only enables us to make $\mu$-SQUIDs capable of operating at fields as high as 4T
independent of the field direction, as reported in this paper, but also finds a potential
application for ultra-sensitive motion detection of diamond based
nanomechanical systems\cite{05Etaki}.

The discovery of superconductivity in
diamond\cite{03Ekimov,04Blase,06Nesladek,07Takano} has opened the
possibility to combine outstanding mechanical properties with
superconductivity. Diamond is the archetype of superhard
materials with the highest Young's
modulus\cite{08Dubitskiy,09Willchemphy}. The main interest in such
super-hard superconducting materials comes from their possible
application to high frequency nanomechanical systems. This high
stiffness and the reduced mass of nanomechanical structures made
out of diamond enable GHz vibration frequencies comparable with or
higher than thermal energies at milliKelvin
temperatures\cite{10Gaidarzhy}. This opens the possibility of
studying the quantum regime of such nanomechanical
resonators\cite{11Cleland,11aTeufel1,11bTeufel2}.

Advances in nanofabrication technology have made it possible to
realise complex devices involving nano-electromechanical
components\cite{12Naik,13Knobel}. But in most cases the
nano-mechanical devices are compound/hybrid devices consisting of
a super-hard component with high Young's modulus generating high
frequency resonators, and of a conducting component to reduce the
transmission losses of the system. However, this conductive
component decreases the mechanical rigidity and thus reduces the
overall quality factor of the device. In this context, boron doped
diamond turns out to be an excellent candidate for fabricating
monolithic superconducting circuits involving nanomechanical
systems with very high quality factor. One can then envision the
direct integration of such nanomechanical systems into
superconducting circuits such as SQUIDs\cite{05Etaki} or
superconducting resonators\cite{14regal} for ultra sensitive
motion detection.

\section*{Results and Discussions}
\begin{figure}[!h]\centerline{\includegraphics[width=2in,angle=0]{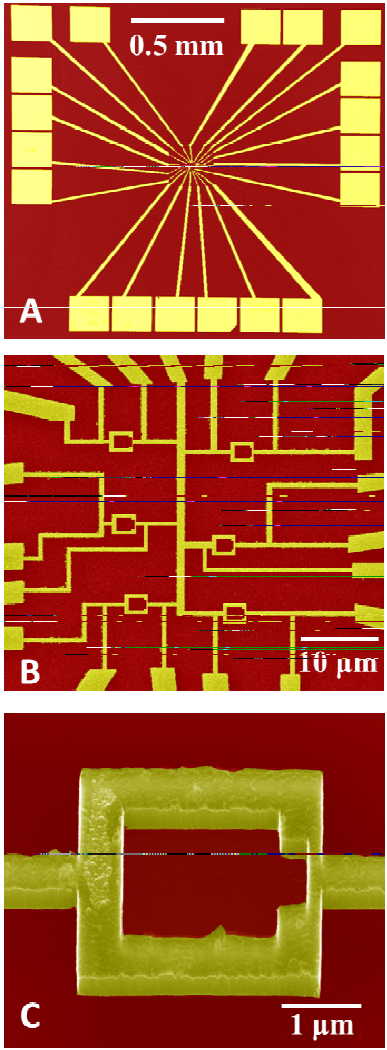}}
\caption{Scanning electron micrograph of the $\mu$-SQUID circuit.
a, Image of the complete diamond circuit showing several pads for
ohmic contacts. The total chip size is 2mm$\times$2mm. b, Close-up
view of a 50$\times$50 $\mu$m$^2$ area in the middle of the
sample. The circuit contains six $\mu$-SQUIDs with different
geometrical characteristics. c, Tilted view of one of the
$\mu$-SQUIDs. The mean loop area of all the $\mu$-SQUIDS is
2.5$\times$2.5 $\mu$m$^2$. The thickness of the diamond film is
300 nm while the arms of the $\mu$-SQUIDs are 500 nm wide. Two
weak links of 170 nm in width and 250 nm in length can be
identified. } \label{1sem}
\end{figure}

As a first step in this direction, we fabricated diamond
$\mu$-SQUIDS patterned from a 300 nm thick superconducting
nano-crystalline diamond film. The boron doped diamond films were
grown by microwave plasma enhanced chemical vapour deposition
(MWPECVD) on seeded silicon {100} wafers with a silica buffer
layer of 500 nm\cite{15Williams}. The thickness was monitored in situ with laser
interferometry. The growth process has been discussed in detail
elsewhere\cite{16Williams}. The diamond thin films were patterned
using standard electron beam lithography. A thin nickel mask (65
nm) was deposited for subsequent highly anisotropic oxygen plasma
etching\cite{17Mandal}. Titanium-platinum-gold was deposited for
the contact pads (figure \ref{1sem}a) and the sample was annealed at
750$^o$ C to ensure good ohmic properties. 

\begin{figure}[h]\centerline{\includegraphics[width=2.5in,angle=0]{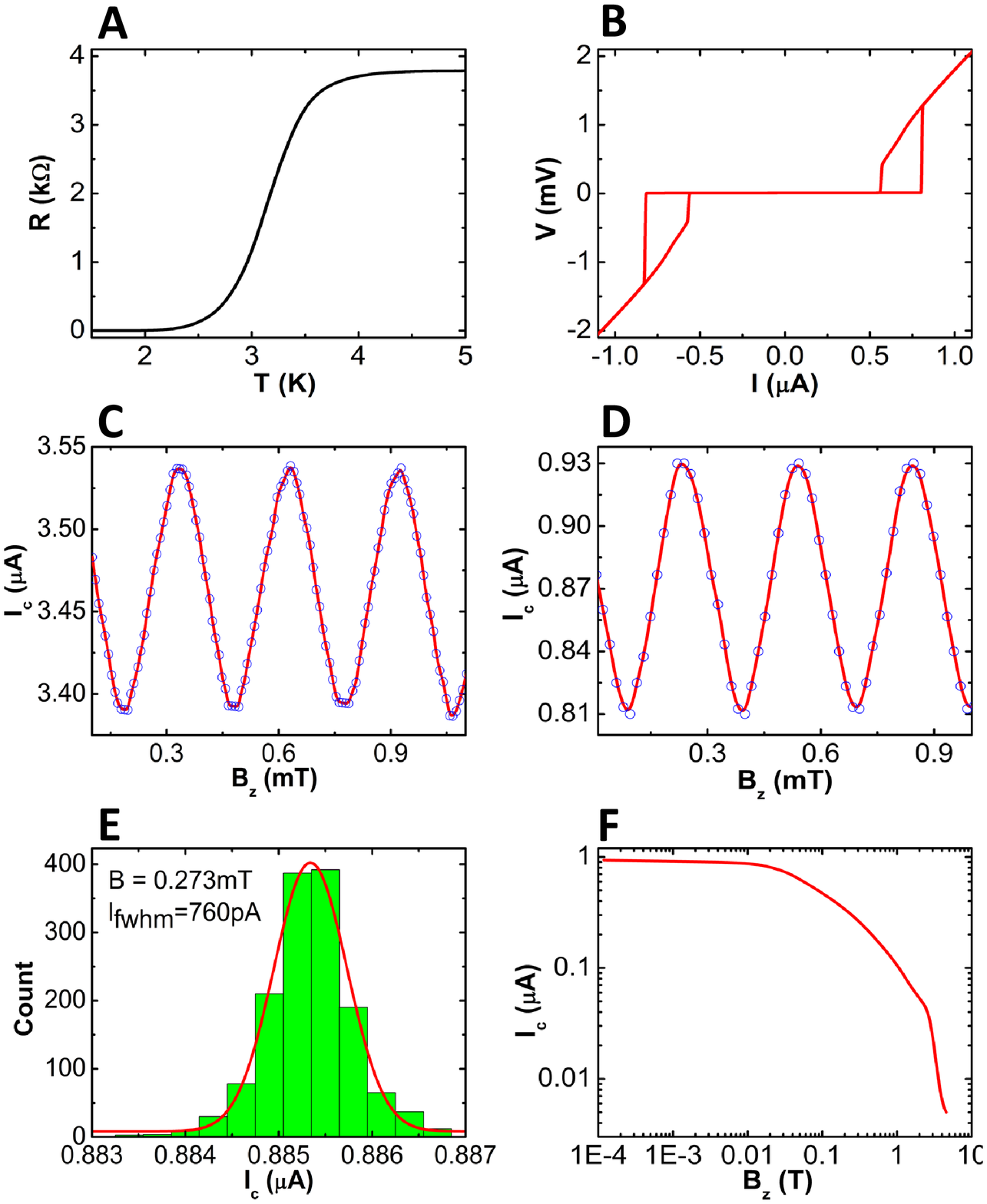}}
\caption{Characteristic features of the diamond $\mu$-SQUID. a,
Superconducting  transition of our $\mu$-SQUID showing a
transition temperature around 3K which is close to the T$_C$ of
the diamond thin film. b, Current voltage characteristics of the
$\mu$-SQUID with 100 nm wide weak link. The I-V curve is
hysteretic with a critical current close to 1 $\mu$A and a
retrapping current of about 0.55 $\mu$A. c, d) Low field
oscillation of the critical current as a function of magnetic
field for a $\mu$-SQUID with a width of a weak link of 170 nm and
100 nm, respectively. The oscillation period in both cases is
around 0.31 mT. e, Histogram of the switching current of the
$\mu$-SQUID with a 100 nm wide weak link. The measurement was done
by recording the switching current repetitively at a fixed
magnetic field of 0.3 mT. f, Field dependence of the critical
current for the same $\mu$-SQUID in perpendicular configuration.}
\label{2lowf}
\end{figure}

We fabricated several similar devices as depicted in figure \ref{1sem}
with various weak link designs of width of 250 nm, 170 nm and 100
nm. These weak links serve as the Josephson junctions in the superconducting loop\cite{17aDayem,17bLikharev}. Here we mainly discuss data measured on a $\mu$-SQUID with an
area of 2.5$\mu$m $\times$ 2.5$\mu$m with two symmetric weak links (100 nm
wide if not stated otherwise). Our low field measurements were
performed in a $^3$He closed cycle refrigerator with a base
temperature of 400 mK, whereas the high field measurements have
been undertaken in a dilution refrigerator at a temperature of 40 mK.
The superconducting critical temperature of the bulk as well as the nanostructured diamond film was about 3K as shown in figure \ref{2lowf}a.

The general characteristics of our $\mu$-SQUIDS have been
summarized in figure \ref{2lowf}. The current voltage (I-V)
characteristic shows a thermal
hysteresis\cite{18aSkocpol,18Courtois} with a critical current of
almost 1$\mu$A and a retrapping current of about 0.55$\mu$A. These
parameters can be changed by tuning the geometrical aspects of the
weak link. To demonstrate the performance of our diamond
$\mu$-SQUIDs we have measured the critical current oscillation as
a function of perpendicular magnetic field. At low magnetic field
we swept the current from a value slightly below the retrapping
current until the switching of the SQUID was detected and
recorded. This was repeated for different magnetic fields. The
results for these experiments are shown in figure \ref{2lowf}c and d. An
oscillation period of approximately 3.1 G was recorded which
corresponds to an effective SQUID surface area of 2.6 $\times$ 2.6
$\mu$m$^2$, in agreement with the geometrical dimensions of the
$\mu$-SQUID. The resulting modulation amplitude was about 15\% for
100 nm wide weak links and around 5\% for a 250 nm width comparable to what is observed for Al and Nb $\mu$-SQUIDs\cite{18aHasselbach}. To probe
the sensitivity of our $\mu$-SQUID we repeated current switching
measurements at a fixed magnetic field. Taking the full width at
half maximum and taking into account the sampling frequency we
obtained a sensitivity of $4 \times 10^{-5}\phi_0/\sqrt{Hz}$,
where $\phi_0$ is the superconducting flux quantum. This sensitivity is comparable to similarly designed $\mu$-SQUIDS
from frequently employed superconductors such as niobium or aluminium\cite{19Hasselbach} but a careful optimization of the SQUID design as well
as material can lead to much higher sensitivities\cite{19aVoss}. At present the sensitivity is only
limited by the electronic measurement set-up or external noise as
the histograms are still symmetric and not limited by quantum
fluctuations even at 40 mK. The main advantage of the diamond
system is that superconductivity persists up to very high magnetic
field (in our case more than 4 T). This is shown by the field
dependence of the critical current as depicted in figure \ref{2lowf}f.

Various types of $\mu$-SQUIDs have been reported in the
literature, which are extremely sensitive magnetic flux detectors and
currently used for scanning SQUID
microscopy\cite{01Cleziou,19Hasselbach}, magnetization
measurements in mesoscopic systems\cite{19bRabaud,20Bluhm} and in isolated
molecules\cite{21Wernsdorfer}. However, a severe drawback of these
detectors is the narrow field range in which they are operable.
In particular, $\mu$-SQUID devices were operated at fields above 1 T
only when the field is applied perfectly in the plane of the
SQUID\cite{22chen}. In addition, in such a parallel field
configuration, the thickness of the superconducting layer has to
be extremely small (of the order of few nanometers), a feature
which drastically reduces the critical current. On the other hand,
operating \textit{standard} $\mu$-SQUIDs in perpendicular field
reduces severely the operational field range well below 1T. Here
we demonstrate that $\mu$-SQUIDs made from boron doped diamond do
not suffer from such limitations and can be operated in magnetic
fields as high as 4 T even when applied in perpendicular
configuration. This is more than a six-fold increase on the
present state of the art\cite{23Finkler}.

\begin{figure}[h]\centerline{\includegraphics[width=2.5in,angle=0]{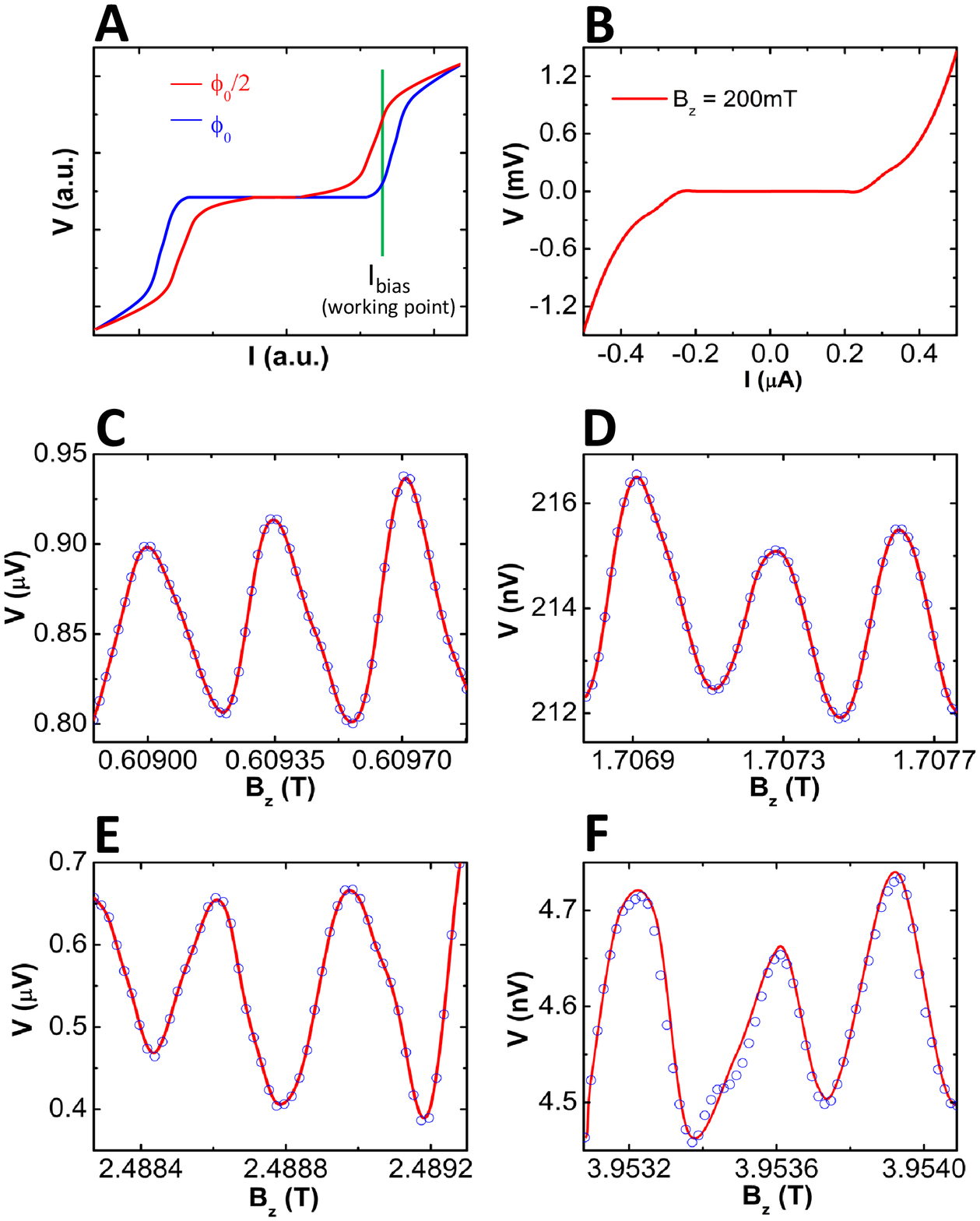}}
\caption{SQUID characteristics in perpendicular field
configuration. a, The schematic of two I-V curves in the
non-hysteretic regime of the device for a applied magnetic flux of
$\phi_0$ and $\phi_0/2$. By current biasing the device at the
green line we can record the oscillation in the output voltage as
the applied magnetic field is swept. b, I-V characteristic when
the SQUID is in the non-hysteretic regime. c-f, SQUID oscillations
at various field ranges (C$\sim$0.6T, D$\sim$1.7T, E$\sim$2.5T,
F$\sim$4T). The average oscillation period is 3.2G, consistent to
what is observed at low magnetic field. The contrasts were not
optimized in these experiments and hence do not follow the field
dependence of the critical current.} \label{3highf}
\end{figure}

In figure \ref{3highf}c-f we have shown the characteristic SQUID oscillations at various
fields up to 4 T. The oscillations are not perfectly periodic due
to the fact that we had to use the z-coil (perpendicular to SQUID
plane) for both, to apply the steady magnetic field (up to 4 T)
and also to probe the SQUID oscillations (of the order of 0.3 mT).
The field resolution is hence limited by the resolution of the
magnet power supply ($\sim$ 0.1 mT). This could be improved by
adding an additional feed line next to the $\mu$-SQUID to probe
the SQUID oscillations independently\cite{05Etaki}.

\begin{figure}[h]\centerline{\includegraphics[width=3in,angle=0]{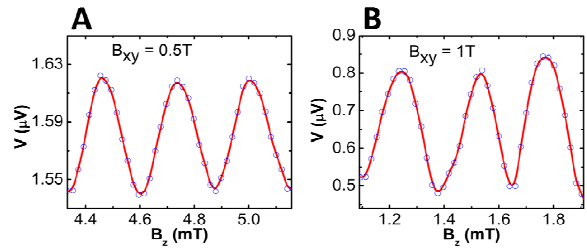}}
\caption{SQUID characteristics in parallel field configuration. a
and b, The voltage oscillations of the SQUID when a constant
magnetic field is applied parallel to the SQUID plane and
perpendicular to both the weak links. To detect the oscillation a
small probing field is applied perpendicular to the SQUID plane.}
\label{4inplane}
\end{figure}

In order to demonstrate the insensitivity of our SQUID to the
applied field direction we have also measured SQUID oscillations
by applying a constant field with a vector magnet in the x-y plane
(parallel to SQUID plane). In this configuration the SQUID
oscillations were probed using a small magnetic field of a few
milli Teslas in the z-coil. The voltage oscillations for a
parallel field of 0.5 and 1 T, the maximal field achievable with
our vector coil, are shown in figure \ref{4inplane}. Let us also
emphasize that these oscillations were obtained for 300 nm thick
superconducting boron doped diamond layers. In this case the
aspect ratio is very favorable for vortex penetration, and in
standard superconductors, current oscillations are usually not
observed in $\mu$-SQUIDS for layers thicker than a few nanometers.
The fact that we are able to observe SQUID oscillations for such
geometries may originate from the granularity of the
material\cite{24Dahlem}, which favors vortex pinning.

Let us mention that the resulting SQUID oscillations at high magnetic
field have not been optimized at each field and for this reason the oscillation
amplitude does not follow the field strength. Here we simply demonstrate the
proof of principle that a diamond SQUID can be operated at high magnetic
field and this independent on the field direction. In order to optimize
the sensitivity of the diamond SQUID it would be extremely interesting to
employ single crystal diamond where higher critical currents\cite{25Watanabe}
as well as higher critical fields ($\sim$10T)\cite{07Takano} can be achieved.

SQUID oscillations surviving at high magnetic fields, independent
of field orientation, open the possibility to exploit these
exceptional properties for HF magnetic surface probe
techniques\cite{02Kirtley}. Combining these exceptional superconducting
and outstanding mechanical properties makes this monolithic system
a highly promising tool, in particular for the detection of
quantum motion in a diamond based nanomechanical oscillator when
inserted within one arm of a diamond SQUID\cite{05Etaki}.

\section*{Methods}
\subsection*{Film Deposition}
Prior to deposition, the wafers were cleaned by standard RCA SC1 solution
and rinsed with deionized water in an ultrasonic bath. Immediately
after rinsing the wafers were immersed in a colloid of
mono-disperse diamond nano-particles known to have a mean size of
6 nm, and agitated by ultrasound for 30 min. Following this seeding
process, the wafers were rinsed again with deionized water, blown
dry in nitrogen and immediately placed inside the growth chamber
which was pumped down to a vacuum lower than 10$^{-6}$ mbar. The
chamber was then purged with hydrogen gas and a plasma ignited
with 4\% methane and 6500 ppm of tri-methylboron diluted by
hydrogen (better than 99.9999999\% pure). The pressure was ramped
to 50 mbar and the temperature was 700$^o$ C as measured by
optical pyrometry.

\subsection*{Electronic Measurements}
The SQUID measurements have been performed by
current biasing the squid loop via a thermostable 1 M$\Omega$
resistor using a 16 bit NI-USB-6229 DAC. At low magnetic field,
the I-V characteristic is hysteretic as seen in figure \ref{2lowf}b and
the SQUID oscillation is measured by recording the critical
current as a function of magnetic field. The critical current is
determined when a voltage drop is generated across the SQUID due
to its transition from the superconducting to the normal state.
The voltage is amplified with a NF-LI75a low-noise voltage
amplifier and recorded via a Keithley 2000 Multimeter. The SQUID
oscillations shown in figure \ref{2lowf} are obtained with a single I-V
curve at a fixed magnetic field. No averaging has been done.

Applying a large magnetic field (B$>$500 mT) to a superconductor reduces the
critical current. When the reduction of the critical current is
below the value of the retrapping current, the voltage-current
characteristics become non-hysteretic as shown in figure \ref{3highf}b. In this case, an ac measurement technique can be used to
measure the SQUID oscillations. The SQUID is biased
with a square wave signal by mixing a sinusoidal voltage
(amplitude of 1 V and a frequency of 11.7 Hz) with the dc voltage
of the NI-DAC. The voltage generated across the SQUID is amplified
with a home-made ultra-low noise voltage amplifier
(0.5nV/$\sqrt{Hz}$) and measured with a Lock-In amplifier (Signal
recovery 7265). An I-V curve measured with this technique is
displayed in figure \ref{3highf}b. By adjusting the DC voltage of the
DAC, the bias current I$_B$ can then be set at the working point of
the SQUID (see figure \ref{3highf}A). Sweeping the magnetic field at fixed
bias current results in voltage oscillations as shown in figures
\ref{3highf}c-f and \ref{4inplane}a,b.

\section*{Acknowledgements}
We would like to acknowledge technical assistance from the Nanofab
team of the Institut N\'eel, in particular B. Fernandez. We also acknowledge valuable
discussions with Wolfgang Wernsdorfer, Klaus Hasselbach and
Vincent Bouchiat and technical assistance from Y. Baines. This work has been supported by the French
National Agency (ANR) in the frame of its program in ``Nanosciences
and Nanotechnologies'' (SUPERNEMS Project no. ANR-08-NANO-033).
This work was partially supported by the Fraunhofer Attract award
``Hybrid HF-MEMS Filters for GHz-Communication and capillary MEMS
systems for chemical and bio-chemical Sensing - COMBIO''.

\end{document}